\documentclass[conference]{IEEEtran}
\IEEEoverridecommandlockouts
\usepackage{cite, balance}
\usepackage{amsmath,amssymb,amsfonts}
\usepackage{algorithmic}
\usepackage{algorithm2e}
\usepackage{graphicx}
\usepackage{textcomp}
\usepackage{xcolor}
\usepackage{url}
\usepackage{array}
\usepackage{multirow}
\usepackage{amssymb}
\usepackage{comment}

\def\BibTeX{{\rm B\kern-.05em{\sc i\kern-.025em b}\kern-.08em
    T\kern-.1667em\lower.7ex\hbox{E}\kern-.125emX}}

\def \toolname {AndroEvolve}
\def \oldtoolname {CocciEvolve}

\definecolor{red}{HTML}{9B0000}
\definecolor{lightred}{HTML}{FF5131}
\definecolor{green}{HTML}{006400}
\definecolor{lightgreen}{HTML}{9CFF57}
\definecolor{purple}{HTML}{7200CA}
\definecolor{verylightgrey}{HTML}{F1F1F1}

\definecolor{diffstart}{named}{blue}%{Grey}
\definecolor{diffincl}{named}{green}
\definecolor{diffrem}{named}{red}

\newcommand{\extrabold}{\bfseries}
\usepackage{multirow}
\usepackage{listings}
  \lstdefinelanguage{diff}{
	basicstyle=\ttfamily\extrabold\scriptsize,
	morecomment=[f][\color{diffstart}]{@},
	morecomment=[f][\color{diffincl}]{+},
	morecomment=[f][\color{diffrem}]{-},
        keepspaces=true,
	identifierstyle=\color{black},
  }

\usepackage[shortlabels]{enumitem}

\usepackage{subcaption}
\usepackage{ifthen}
\newboolean{showcomments}
\setboolean{showcomments}{true}
\ifthenelse{\boolean{showcomments}}
{ \newcommand{\mynote}[2]{\textcolor{red}{
			\fbox{\bfseries\sffamily\scriptsize#1}
			{\small$\blacktriangleright$\textsf{\emph{#2}}$\blacktriangleleft$}}}}
{ \newcommand{\mynote}[2]{}}

\ifthenelse{\boolean{showcomments}}
{ \newcommand{\hnote}[2]{\textcolor{blue}{
			\fbox{\bfseries\sffamily\scriptsize#1}
			{\small$\blacktriangleright$\textsf{\emph{#2}}$\blacktriangleleft$}}}}
{ \newcommand{\hnote}[2]{}}

\newcommand{\jlx}[1]{\mynote{JLX}{#1}}

\begin{document}

\author{
\IEEEauthorblockN{
Stefanus A. Haryono\IEEEauthorrefmark{1},
Ferdian Thung\IEEEauthorrefmark{1},
David Lo\IEEEauthorrefmark{1},
Lingxiao Jiang\IEEEauthorrefmark{1},
Julia Lawall\IEEEauthorrefmark{3},
Hong Jin Kang\IEEEauthorrefmark{1},\\
Lucas Serrano\IEEEauthorrefmark{2}, and
Gilles Muller\IEEEauthorrefmark{3},
}

\IEEEauthorblockA{\IEEEauthorrefmark{1}School of Information Systems, Singapore Management University, Singapore\\
\{stefanusah,ferdianthung,davidlo,hjkang.2018,lxjiang\}@smu.edu.sg}
\IEEEauthorblockA{\IEEEauthorrefmark{2}Sorbonne University/Inria/LIP6, France\\
Lucas.Serrano@lip6.fr}
\IEEEauthorblockA{\IEEEauthorrefmark{3}Inria, France\\
\{Gilles.Muller,Julia.Lawall\}@inria.fr}
}

% \author{\IEEEauthorblockN{Stefanus Agus Haryono}
% \IEEEauthorblockA{School of Information System\\
% Singapore Management University\\Singapore\\
% stefanusah@smu.edu.sg}
% \and
% \IEEEauthorblockN{Ferdian Thung}
% \IEEEauthorblockA{School of Information System\\
% Singapore Management University\\Singapore\\
% ferdianthung@smu.edu.sg}
% \and
% \IEEEauthorblockN{David Lo}
% \IEEEauthorblockA{School of Information System\\
% Singapore Management University\\Singapore\\
% davidlo@smu.edu.sg}
% \and
% \IEEEauthorblockN{Lingxiao Jiang}
% \IEEEauthorblockA{School of Information System\\
% Singapore Management University\\Singapore\\
% lxjiang@smu.edu.sg}
% \and
% \IEEEauthorblockN{Julia Lawall}
% \IEEEauthorblockA{Inria\\France\\
% Julia.Lawall@inria.fr}
% \and
% \IEEEauthorblockN{Hong Jin Kang}
% \IEEEauthorblockA{School of Information System\\
% Singapore Management University\\Singapore\\
% hjkang.2018@smu.edu.sg}
% \and
% \IEEEauthorblockN{Lucas Serrano}
% \IEEEauthorblockA{Sorbonne University/Inria/LIP6\\France\\
% Lucas.Serrano@lip6.fr}
% \and
% \IEEEauthorblockN{Gilles Muller}
% \IEEEauthorblockA{Inria\\France\\
% Gilles.Muller@inria.fr}
% }

\captionsetup{font=footnotesize,belowskip=1pt,aboveskip=1.0pt}

\pagestyle{plain}
\pagenumbering{arabic}
\title{AndroEvolve: Automated Android API Update with Data Flow Analysis and Variable Denormalization}

% \author{Stefanus Agus Haryono$^*$, Ferdian Thung$^*$, Kang Hong Jin$^*$, Lucas Serrano$^{\dagger}$, Gilles Muller$^{\ddagger}$, Julia Lawall$^{\ddagger}$, David Lo$^*$, Lingxiao Jiang$^*$}

% \author{Stefanus A. Haryono$^*$, Ferdian Thung$^*$, Hong Jin Kang$^*$, Lucas Serrano$^{\dagger}$, Gilles Muller$^{\ddagger}$,\ \ \ \ \ \ \ \ \ \ \ \ \ \ \ \ \ \ \  Julia Lawall$^{\ddagger}$, David Lo$^*$, Lingxiao Jiang$^*$}
% \affiliation{%
%   \institution{*School of Information Systems, Singapore Management University, Singapore}
% }
% \email{{stefanusah,ferdianthung,hjkang.2018,davidlo,lxjiang}@smu.edu.sg}
% % \email{hjkang.2018@phdcs.smu.edu.sg}
% \affiliation{%
%   \institution{$^{\dagger}$Sorbonne University/Inria/LIP6, France}
% }
% \email{Lucas.Serrano@lip6.fr}
% \affiliation{%
%   \institution{$^{\ddagger}$Inria, France}
% }
% \email{{Gilles.Muller,Julia.Lawall}@inria.fr}  

\maketitle

\begin{abstract}
The Android operating system is frequently updated, with each version bringing a new set of APIs. New versions may involve API deprecation; Android apps using deprecated APIs need to be updated to ensure the apps' compatibility with old and new versions of Android. Updating deprecated APIs is a time-consuming endeavor. Hence, automating the updates of Android APIs can be beneficial for developers. \oldtoolname{} is the state-of-the-art approach for this automation. However, it has several limitations, including its inability to resolve out-of-method-boundary variables and the low code readability of its update due to the addition of temporary variables. In an attempt to further improve the performance of automated Android API update, we propose an approach named \toolname{}, which addresses the limitations of \oldtoolname{} through the addition of data flow analysis and variable name denormalization. Data flow analysis enables \toolname{} to resolve the value of any variable within the file scope. Variable name denormalization replaces temporary variables that may present in the \oldtoolname{} update with appropriate values in the target file. We have evaluated the performance of \toolname{} and the readability of its updates on 360 target files. \toolname{} produces 26.90\% more instances of correct updates compared to \oldtoolname{}. Moreover, our manual and automated evaluation shows that \toolname{} updates are more readable than \oldtoolname{} updates.

\end{abstract}

\begin{IEEEkeywords}
Program transformation, Android, data flow analysis, readability, API deprecation, API update
\end{IEEEkeywords}

\section{Introduction}

Android is currently one of the most prominent operating systems (OS) due to the vast amount of its users. 
% Being a popular platform, Android OS is frequently updated. 
Android OS is frequently updated to add new features or to fix bugs. With each new version, changes and modifications in its APIs are inevitable. Changes to Android APIs may deprecate older versions and render them unusable in the newer version of the OS. To prevent errors caused by such API deprecation, developers need to constantly update deprecated-API usages in their code, while still maintaining backward compatibility with older Android versions. This problem, termed Android fragmentation~\cite{han2012understanding, wei2016taming}, is a common occurrence. Aside from being cumbersome and time-consuming to mitigate, Android fragmentation also introduces security risks~\cite{fragmentationsecurity}.

Due to the nature of Android fragmentation, updating usages of deprecated Android APIs has become a priority. To help developers, several studies have proposed automatic approaches for updating Android API usages~\cite{fazzini2019automated, coccievolve}. AppEvolve is a recent approach~\cite{fazzini2019automated}. It uses both before- and after-update code examples to learn the update automatically. However, while it is able to provide an applicable update for some examples, 
it was found to have several weaknesses and
a replication study by Thung et al.\cite{thung2020automated} demonstrated that AppEvolve works correctly only when the target file to be updated has a very similar syntax to the code example. %They further demonstrate these weaknesses through several experiments.

More recently, Haryono et al.\cite{coccievolve} presented a new tool for automatic Android API update called \oldtoolname{}
built on Coccinelle4J~\cite{lawall2018coccinelle}.
\oldtoolname{} shows better performance than AppEvolve on 112 target files. 
% \ft{We should say this approach beats AppEvolve?} \sa{added a sentence about this}
The main improvements that \oldtoolname{} provides are:
%an easily readable \jlx{remove the phrase, because, if it's "easily" already, then why our new tool still needs to improve its readability?}
generated update-scripts in the Semantic Patch Language (SmPL)~\cite{lawall2018coccinelle}, updates generated by using only a single after-update example, and its capability to update multiple instances of deprecated API invocations within a single file. 
However, the code example used in \oldtoolname{} must be in the form of an {\tt if} statement containing the updated API in the ``then'' statement and the old API in the ``else'' statement, or vice versa. A sample of such an after-update code example can be seen in Figure~\ref{fig:example_update_code}.
% \ft{I think after-update code example is better than updated example code? If you agree, let's use this term throughout the paper.} \sa{i think after-update code example is better. Better use that term throughout the paper}

\begin{figure}[h]
	\centering
	\scriptsize{
\begin{lstlisting}[language=java,numbers=none,sensitive=true,columns=flexible,basicstyle=\ttfamily]
if (android.os.Build.VERSION.SDK_INT >= 
        android.os.Build.VERSION_CODES.M) {
    minutes = picker.getMinute();
} else {
    minutes = picker.getCurrentMinute();
}
\end{lstlisting}
        \setlength{\belowcaptionskip}{7pt}
		\caption{An example of after-update code for {\tt getCurrentMinute()} API}\label{fig:example_update_code}
	}
\end{figure}

% \ft{I think this paragraph is no longer necessary.} \sa{just delete the whole paragraph of itemized item?}\ft{Yes, I think so. We should take more about our approach instead.}However, despite its improvements, \oldtoolname{} has some shortcomings:
% \begin{itemize}[nosep,leftmargin=*]
%     \item \oldtoolname\ is incapable of resolving values that are used as updated API arguments;
%     %\ft{What are these values? We probably should describe more. I'm not sure whether people can understand it. Maybe by an example? Would be good also to explain why \toolname{} cannot resolve them?}\sa{Will add further explanation of this problems in a new paragraph}    \ft{Ok thanks}
%     \item \oldtoolname\ adds temporary variables in its process that are undesirable and may reduce code readability;
%     \item \oldtoolname\ only provides generic variable name that makes its result hard to understand.\ft{Are we still handling this?} \sa{I think we are handling this by removing the temporary variables. But if it is not correct, maybe can remove this point}\ft{I see, we can explain this in more detail in another section.} \sa{I think since this paragraph is going to be deleted, it does not matter?}
% \end{itemize}

The major problem of CocciEvolve is its inability to resolve all the values used as API arguments.
%that may come from outside of method boundaries. 
These values in API method invocations in a method body can be expressions in various forms, such as literal expressions, name expressions, field access expressions, method invocations, and object creations. 
%For literal expressions (e.g. an integer, String) CocciEvolve can work correctly by directly copying them. However, if the expressions are in a form other than literal expressions, CocciEvolve fails to produce a working update. 
When the expressions refer to a variable defined outside of the method boundary, CocciEvolve fails to resolve the variable or produce a working update.
Another problem in \oldtoolname{} is about the readability of its update results. During the update process, CocciEvolve introduces temporary variables that refer to other variables in the target file. The temporary variables are used to ease the transformation process in \oldtoolname{}, but %However, these temporary variables are not deleted and 
they clutter the update results.
%This problem affects the readability of the updated code produced by \oldtoolname{}.

In this work, we propose \toolname{}, an improved automated Android API usages update tool that addresses the limitations of \oldtoolname{}, with two new features:
%to address those limitations: 
data flow analysis and variable name denormalization.
During the update-script creation, data flow analysis is used to resolve the values used as API arguments, including all variables that are defined outside of the current method containing deprecated API invocations to be updated. Definitions of such out-of-method-boundary variables are located
%based on the program control flow 
and used to replace the variables in the API invocations.
For brevity, we refer to such variables as out-of-method variables in the rest of the paper.
%Data flow analysis allows for more flexibility in the code that is used as the after-update example.

\begin{comment}
\jlx{this paragraph doesn't seem to be useful, although it explains where comes the temporary variables; may keep the explanation short if not the explanation doesn't show the advantage of our new tool.
Modified.}
Conforming with the approach taken by \oldtoolname{}, source file normalization is also used in \toolname{} as a {\em pre-processing step}. Source file normalization mitigates the problem of failed update that is caused due to minor syntactic differences between the after-update example and the target file. Through source file normalization, the API invocation and its arguments are converted into normalized form to ease the update process by introducing temporary variables. Each temporary variable is an assignment which contains the reference to the arguments and class or object used in the API invocation.
\end{comment}
 
Variable name denormalization is added to improve the readability of the updated code as a {\em post-processing step}.
As temporary variables can be introduced by \oldtoolname{} to normalize syntactic differences between update examples and the target file to ease the code update process, the readability of the updated code may be decreased with more temporary variables used. 
Our denormalization refers to a process that converts the code normalized to use temporary variables back to their original form {\em after} updates have been performed
%This process removes the temporary variables introduced by the source file normalization process and replaces their usages with the original values referred by these temporary variable definitions. Removing these temporary variables 
and thus improves the readability of the produced updates.

% \sa{Changed into the 2 paragraphs above}
% \ft{It might be better to explain this in a paragraph instead of bullet points. We can explain these along with explaining how \toolname{} works}
% extension and improved version of \oldtoolname{}\ft{This may give an indication that \oldtoolname{} is created by the same authors of this paper} \sa{what should we do or how should we said this then?}\ft{We can probably say \toolname{} addresses the limitations of \oldtoolname{} directly. I think saying it a version of \oldtoolname{} is the giveaway that the tool is created with the same authors. I guess we are okay if we do not say that.} that address their limitations. Improvements that we bring with \toolname{} includes:
% \begin{itemize}[nosep,leftmargin=*]
%     \item \toolname\ adds the capability of data flow analysis to resolve out of method boundaries values in the update script creation. This allows for more flexibility in the syntax of the codes that are used as the update example.
%     \item \toolname\ adds variable name denormalization\ft{Need to describe what is denormalization}\sa{describe it where? I think my description will be: Denormalization refers to the process of reverting back the code that was normalized into the temporary variables into their original form in the target code.} to remove the temporary variables introduced by \oldtoolname{} update and substitute them with their original value. Removing these temporary variables increase the readability of the produced updates.
% \end{itemize}

To evaluate the performance of \toolname{}, we have conducted an %extensive 
experiment using a dataset of 360 target files containing 20 different Android APIs.
% \ft{the dataset has 20 though?} \sa{But only 16 APIs are used. If that is the case, should we mention 20 and explain later why the other 5 APIs does not work?}\ft{Yes we should still include the APIs in the dataset if we run our approach on them.}\ft{Looking at Table~\ref{table:data_statistic}, some APIs have no target code. I thought we should have at least a target code from AppEvolve dataset?} \sa{we can easily add them since they won't work anyway. So let's say we want to add 3 data for each of the non working API? I don't think we are supposed the use the target code from AppEvolve dataset since our dataset is based on the CocciEvolve one (ICPC paper). Should I just add 3 data for each of the non-working 4 APIs? Or do you prefer any other number? It is simple to look for the data}\ft{Any number is okay as long as it's not zero. We can add from AppEvolve also if we want as the target code is the same.} \sa{will add 2 for each of the APIs. However, I don't think we should add target for the shouldOverrideURLLoading API do we? Since that API does not have any after-update example}
We have compared the performance of \toolname{} against \oldtoolname{} by counting the number of successful updates produced by each tool. \toolname{} has produced 26.9\% more successful updates.
For readability, we have compared the update results of \toolname{} and \oldtoolname{} through both an automated measurement by using a popular readability scoring tool~\cite{readabilitymodel} and a manual measurement by asking the opinions of two experienced Android engineers. The measurements highlight that \toolname{} produces updates that are about 50\% and 83\% more readable than \oldtoolname{} with respect to the automated and manual measurements
%have better readability compared to those produced using \oldtoolname{}; the automated and manual assessment show that AndroEvolve outperforms CocciEvolve by 49.89\% and 82.94\% 
respectively.

The main contributions of our work are as follows:
\begin{enumerate}[nosep,leftmargin=*]
    \item We propose \toolname{}, a tool that addresses the limitations of \oldtoolname{}. \toolname{} adds data flow analysis to resolve out-of-method variables and introduces variable name denormalization to increase update readability.
%\jlx{too much repetitive description below with the previous paragraph; may shorten somehow; Edited.}
%    \item We evaluate the update performance of \toolname{} by counting the number of successful updates produced for a dataset containing 360 target files involving 20 Android APIs and shows that it outperforms \oldtoolname{} by 26.90\% in terms of the number of successful updates produced.
%    \item We measure the readability of updated code using both a manual and an automated approach. Both report higher readability for the updated code produced by \toolname{} when compared to \oldtoolname{}; The automated and manual assessment of \toolname{} show that it outperforms \oldtoolname{} by 49.89\% and 82.94\% respectively.
    \item We evaluate \toolname{} on a dataset containing 360 target files involving 20 Android APIs and show that it outperforms \oldtoolname{} in terms of both update success rate and the readability of the updates.
\end{enumerate}

\smallskip
The rest of this paper is organized as follows. Section~\ref{sec:prelim} provides preliminaries on CocciEvolve, data flow analysis, and code readability. Section~\ref{sec:motivating_examples} provides the motivating examples that show the problems present in \oldtoolname{}. Section~\ref{sec:approach} discusses our approach in creating \toolname{} as the upgraded version of \oldtoolname{}. Section~\ref{sec:exp} provides the details on the experiments and its results. Section~\ref{sec:discuss} discusses the limitations of our work. Lastly, Section~\ref{sec:conclusion} concludes our work and provides a discussion of future plans.

\section{Preliminaries}\label{sec:prelim}
{\bf CocciEvolve}~\cite{coccievolve} is the state-of-the-art tool on automatic Android API usage update. It distinguishes itself by only requiring a single after-update example, providing a readable update-script, and introducing code normalization that
tolerates some syntactic differences during code updates.
%enables a more robust update. %These features were proposed by \oldtoolname{} to achieve a better performance in automatic Android API update.
%\oldtoolname{} is a recent tool on automatic Android API usage update that has a higher performance than the previous state-of-the-art tool, AppEvolve. Compared to AppEvolve, \oldtoolname{} distinguished itself by only using a single after-update example, providing a readable and understandable update script and process through Coccinelle4J utilization, and code normalization that enables more robust update. These features were proposed by \oldtoolname{} to achieve a better performance in automatic Android API update

%HJ : I replaced kang2019automating with kang2019semantic. They're actually rhe same paper, one published at ECOOP and one is an Inria report. Let's stick to the ECOOP one (peer-reviewed).
%SA : Thank you very much for the clarification!
    Firstly, using only a single after-update example,
    \oldtoolname{} becomes applicable to more cases than its predecessor AppEvolve~\cite{fazzini2019automated} that requires both before- and after-update example.
    %\oldtoolname{} increases the number of usable examples, when compared to AppEvolve~\cite{fazzini2019automated} (i.e., the previous state-of-the-art tool) that requires before-update and after-update examples. 
    Secondly, a readable update-script is achieved through the use of Coccinelle4J~\cite{kang2019semantic}. Coccinelle4J is a program matching and transformation tool for Java language, ported from Coccinelle~\cite{lawall2018coccinelle,padioleau2008documenting}.
    Coccinelle4J describes its transformation using a patch written in Semantic Patch Language (SmPL), which has similar syntax with \textit{diff}. Having a patch that describes the program transformation helps the developers to better understand the updates and transformations applied.
    Finally, by doing a normalization on both the after-update example and the target file, \oldtoolname{} minimizes the syntactic differences that may cause a failed update. Syntactic differences occur when the after-update and the target file API invocation arguments are expressed in different syntax. This was one of the main limitations of AppEvolve, as shown in a replication study by Thung et al.~\cite{thung2020automated}.
%First, by using only a single after-update example, \oldtoolname{} increase the amount of usable after-update example compared to AppEvolve that requires both before-update and after-update example. The second improvement is through the use of Coccinelle4J. Coccinelle4J~\cite{kang2019semantic} is a program matching and transformation tool for Java language, that is ported from Coccinelle\cite{lawall2018coccinelle,padioleau2008documenting}. Coccinelle4J describe its transformation using semantic patch written in semantic patch language, which has similar syntax with \textit{diff}. Having semantic patch that describes the program transformation helps the developers to get a better understanding on the updates and transformations applied. Finally, by doing a normalization on both the after-update code and the target code, \oldtoolname{} minimize the problem of syntax mismatching that caused a failed update. Syntax mismatch occurs when the after-update and the target code have the same semantic that is expressed in different syntax. This was one of the main problem in AppEvolve, as shown in the replication work of Thung et al.~\cite{thung2020automated}.

\vspace{0.2cm}\noindent{\bf Data flow analysis}, such as def-use analysis, is an analysis of the data within the code based on the control flow paths taken by the program. For each given expression at a point inside a program, data flow analysis 
%combines all the collected information and fact (previous expressions, definitions, etc.) to 
can determine the value of the expression.
Sample uses of data flow analysis are dead code elimination, variable value prediction, and program slicing.
For our work, data flow analysis is mainly used to determine the values of variables used in the arguments for API invocations and to conduct program slicing.
% Preliminaries about data flow analysis

\vspace{0.2cm}\noindent{\bf Code readability} is a measure of how easy it is to read a piece of code. It is an important code feature that developers look out for, especially for code that needs to be maintained for a long run, or code that is touched by multiple developers. Having readable code makes it easier for developers to understand and modify the code. Studies on code readability have been conducted extensively~\cite{readabilitymodel, readabilityempirical, busereadability, textualreadability}.
These studies define the metrics and features that are considered as important factors in determining code readability. These features include structural features (e.g. numbers of lines of code, length of each line of code, etc.) and textual features (e.g. name of variables, consistency between comments and variable name, etc.).
%Readability is a measure of how easy it is to read a text. Code readability is a measure of readability of a code. Code readability is critical aspect for developers, especially for code that needs to be maintained for a long time, or code that is worked on by multiple developers. Having a readable code makes it easier for developers to understand and work on the code. Studies on code readability have been conducted extensively on multiple occasions \cite{readabilitymodel, readabilityempirical, busereadability, textualreadability}. These studies define the metrics and features that are considered as factors in determining the readability of a code. These features include structural features (e.g. line of code, length of code, etc) and textual features (e.g. name of variables, coherence between variables, etc).

\section{Motivating Example}\label{sec:motivating_examples}
The two major limitations of \oldtoolname{} are its inability to resolve variables defined outside of the current method containing invocations to deprecated APIs (so called {\it out-of-method variables}) and the presence of temporary variables in the updated code. 
%In this section, we show examples of such problems.

%JLX: remove "-boundary" from the term.
\textbf{Out-of-method variables.} An after-update example for the {\tt requestAudioFocus(...)} API is shown in Figure~\ref{fig:out_of_method_boundaries}. In this example, the deprecated API that is going to be updated from the target file is the {\tt requestAudioFocus(...)} in line 77. The updated API and its argument is shown in line 74--75. In this example, the method invocation argument for the updated API is the {\tt request} object, shown in line 75. This object needs to be defined through a method invocation of {\tt audioFocusRequestOreo.getAudioFocus-
Request()} (line 74).
% \jlx{is the global variable audioFocusRequestOreo there already in the update example? If yes, then why can't just follow the example considering the diff? Why must use data-flow analysis?}
% \sa{The global variable exist in the after-update example and not exist in the target files to be updated. \oldtoolname{} and \toolname{} does not work based on the whole file diff, but rather works on the code slice that contains the API invocation and its arguments. An example of this code slice can be seen in Figure~\ref{fig:example_update_code}. Based on Figure~\ref{fig:out_of_method_boundaries}, the {\tt audioFocusRequestOreo} object is located outside of the method boundary of the API invocation, thus is not a part of the after-update code slice that will be used in the update script creation. This is why we need the data flow analysis to resolve all the values that are actually needed in this code slice.}
% \jlx{Yes, like Julia mentioned, probably the descriptions need to indicate which lines of code of this example are in after-update (and not in the target files), which lines of code are in the target file, and learning from after-update examples can't tell how to define the out-of-method variables. Same for the example in Fig. 5. And, you may comment out my comments after you address them :-) Thanks!}
% \sa{Thank you! I will try to add these descriptions to make it clearer.}
% This method and its class are defined outside of the {\tt tryToGetAudioFocus} method, as can be seen in line 110. 
The variable {\tt audioFocusRequestOreo} is also defined outside at line 41.
This variable and its definition is a new argument that is not yet defined in the deprecated API invocation, thus they are not present in the target file.
Since CocciEvolve only performs an intra-procedural analysis of update examples and only considers line 74--75 for the update script creation, the variables related to the {\tt request} object (line 41, 110--133) cannot be resolved, hence creating a non-working update.
%This problem limits the syntax of the code that can be used as an update example.

Our proposed solution is the addition of data flow analysis as a variable value resolver. We use data flow analysis on the after-update example's code to find the definitions of the variables used in the API invocation's arguments in a file scope.
%This approach allows \toolname{} to resolve any variable's value in a file scope.
To further improve the functionality of this data flow analysis across files and Java classes, we also copy method and class definitions. If a variable's value is resolved as a method invocation or an object creation, its method or class definition is needed to create a working update. \toolname{} copies such definitions into the updated file, allowing the uses of those method invocations or object creations.

\begin{figure}[t]
	\centering
	\scriptsize{
\begin{lstlisting}[language=java,numbers=none,sensitive=true,columns=flexible,basicstyle=\ttfamily]

41  AudioFocusRequestOreo audioFocusRequestOreo = new 
        AudioFocusRequestOreo();
    ...
67  public void tryToGetAudioFocus() {
68      OnAudioFocusChangeListener listener = this;
69      int result;
70      int type = AudioManager.STREAM_MUSIC;
71      int duration = AudioManager.AUDIOFOCUS_GAIN;
72      if (android.os.Build.VERSION.SDK_INT >= 
                android.os.Build.VERSION_CODES.O) {
74          AudioFocusRequest request = audioFocusRequestOreo.
                    getAudioFocusRequest();
75          result = audioManager.requestAudioFocus(request);
76      } else {
77          result = audioManager.requestAudioFocus(listener,
                type, duration);
78      }
79  }
    ...
110 private class AudioFocusRequestOreo {
111     public AudioFocusRequest getAudioFocusRequest() {
            ...
132     }
133 }
\end{lstlisting}
		\caption{Sample out-of-method argument for an API invocation {\tt requestAudioFocus(...)}}\label{fig:out_of_method_boundaries}
	}
\end{figure}

\textbf{Temporary variables in update results.}
Temporary variables are used in \oldtoolname{} to ease the process of code update. However, these variables remain in the updated code, affecting its readability.
%are not deleted affecting the readability of the update result. 
Furthermore, these variables only refer to other variables that are already in the actual target file. Consider the sample updated code in Figure~\ref{fig:temporary_variables}, there exist two temporary variables, {\tt parameterVariable0} (line 2) and {\tt classNameVariable} (line 3). These variables refer to other parameters of the {\tt setTimeH} method, and can be removed and replaced by their definitions. 
To resolve this problem, our proposed solution is to add variable name denormalization. This denormalization removes the declarations and definitions of the temporary variables,
%that were introduced during the normalization process in the update. This will also 
and replace uses of such temporary variables with the original variables that are referred to by those temporary variables. For example, line 2 in Figure~\ref{fig:temporary_variables} will be removed and {\tt parameterVariable0} in line 5 and 7 will be replaced by {\tt hour}.
%, the original variable that is used as the {\tt setCurrentHour(...)} API invocation argument before the normalization process.

\begin{figure}[h]
	\centering
	\scriptsize{
\begin{lstlisting}[language=java,numbers=none,sensitive=true,columns=flexible,basicstyle=\ttfamily]
1 public void setTimeH(TimePicker tp, int hour) {
2     int parameterVariable0 = hour;
3     TimePicker classNameVariable = tp;
4     if (Build.VERSION.SDK_INT >= Build.VERSION_CODES.M) {
5         classNameVariable.setHour(parameterVariable0);
6     } else {
7         classNameVariable.setCurrentHour(parameterVariable0);
8     }
9 }
\end{lstlisting}
		\caption{An example of temporary variables in the output of CocciEvolve after updating the {\tt setCurrentHour(...)} API invocation.}\label{fig:temporary_variables}
	}
\end{figure}

\section{Approach}\label{sec:approach}
\subsection{\toolname{} Overview}
\begin{figure}[t]
	\centering  
	\captionsetup{belowskip=1.0pt,aboveskip=3.0pt}
	\includegraphics[width=0.85\linewidth]{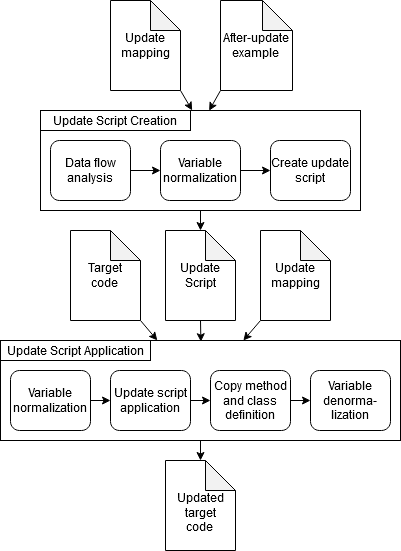}
	\caption{Summary of \toolname{} workflow}
	\label{fig:androevolve_summary}
\end{figure}

The workflow of \toolname{} is comprised of two main functionalities, update-script creation, and update-script application.
Figure~\ref{fig:androevolve_summary} provides a graphical description of this workflow.
Update script creation takes as input the API update mapping and the after-update example of the API to create the update script. The API update mapping defines the mapping between the deprecated API and the updated API. Within the update script creation process, several components are at work. First, data flow analysis is used to resolve any out-of-method variables used by the updated API arguments in the after-update example. This data flow analysis also locates the definitions of any method invocations or object creations used as API arguments. Following the data flow analysis, source file normalization is done on the code block containing the API invocation. Variable normalization introduces temporary variables as replacements for the API invocation arguments to facilitate the update process. 
% the normalization and temporary variables ease and facilitates the update process because they prevent the problem of syntactic differences. Possible explanation can be found in the Preliminaries, second paragraph
Finally, an update script is created based on this modified example code.

This update script, along with the API update mapping and the target file, is the input for the update application process. Within the update application process, there are also several steps that are applied. First, source file normalization is also used in the target file to ease the update process. The update script is then applied to the normalized code. After the update is done, we copy the methods and class definitions that are used by the method invocation or object creations that are used in the updated API arguments. Finally, we apply variable name denormalization to remove temporary variables introduced by the variable normalization and replace their usage with the original expressions used as the API arguments. Details of each functionality are provided below.
%Details of each functionality are provided in the following subsections.

\subsection{Data Flow Analysis}
API method invocations may include arguments that are syntactically different but semantically equivalent.
% API method invocations may include arguments in the form of expressions. 
% Expressions in Java are constructed using smaller program elements such as variables, operators, method invocations, and literal values. This variety of expressions may cause syntactic differences in the expressions used as API invocation arguments that have the same semantics.
Figure~\ref{fig:arguments_form} shows an example of different forms of arguments for {\tt setAudioAttributes} method invocations. In the first example, the argument is first instantiated and assigned into a variable (line 2, 3) before being used as the argument for the first {\tt setAudioAttributes} method invocation (line 10). The {\tt builder} variable (line 3) in this example can be a {\it free} out-of-method variable (only defined in line 2 outside of the method containing lines 9 and not passed as an argument).
%since it is not locally declared nor passed as a parameter to the {\tt setAudioAttributes} API invocation. 
%\jlx{what's "free" above? Added explanation in relation to the term used previously.}
The second example shows a code fragment where a complex expression is directly put as an argument of the second {\tt setAudioAttributes(...)} method invocation (line 26). Contrary to the first example, the argument used in this example is bound (i.e., a variable locally defined or passed in via a method parameter).

\begin{figure}[t]
	\centering
	\scriptsize{
\begin{lstlisting}[language=java,numbers=none,sensitive=true,columns=flexible,basicstyle=\ttfamily]
    // First Example
1   public class AudioPlayer {
2       AudioAttributes.Builder builder = new 
            AudioAttributes.Builder();
3       AudioAttributes attributes = builder.build();
        ...
8       private void setAttributes() {
9           if (android.os.Build.VERSION.SDK_INT >=
                android.os.Build.VERSION_CODES.LOLLIPOP) {
10              mMediaPlayer.setAudioAttributes(attributes);
11          } else {
                ...
            }
        }
    }
    // Second Example
20  public class AudioPlayer {
        ...
24      private void setAttributes() {
25          if (android.os.Build.VERSION.SDK_INT >=
                android.os.Build.VERSION_CODES.LOLLIPOP) {
26              mMediaPlayer.setAudioAttributes(new 
                    AudioAttributes.Builder().build());
27          } else {
                ...
            }
        }
    }
    // Normalized first example
30  public class AudioPlayer {
30      AudioAttributes.Builder builder = new 
            AudioAttributes.Builder();
31      AudioAttributes attributes = builder.build();
        ...
38      private void setAttributes() {
39          if (android.os.Build.VERSION.SDK_INT >=
                android.os.Build.VERSION_CODES.LOLLIPOP) {
40              AudioAttributes parameterVariable0 = attributes;
41              MediaPlayer classNameVariable = mMediaPlayer;
42              classNameVariable.setAudioAttributes(
                    parameterVariable0);
43          } else {
                ...
            }
        }
    }
\end{lstlisting}
		\caption{An example of different forms of argument for {\tt setAudioAttributes} method invocation}\label{fig:arguments_form}
	}
\end{figure}

%\toolname{} creates an update script by extracting the relevant piece of code from the code example. It addresses one of the \oldtoolname{} limitations.
Suppose that \oldtoolname{} is given an after-update example as shown in the first example in Figure~\ref{fig:arguments_form}. Part of the code containing the API invocation and its argument is first normalized, resulting in the normalized first example code (line 30--44). Slice of the normalized code that is used to create the update script is the part contained within the {\tt if} statement, shown in line 40--42.
Using this normalized code slice, \oldtoolname{} will produce an incorrect update, as shown in Figure~\ref{fig:old_coccievolve_result_wrong}. This update script is incorrect since based on the code slice, {\tt newParametervariable0} is only resolved to the bound variable, {\tt attributes} (line 40). This is because \oldtoolname{} cannot resolve the correct value of the expression used as the API invocation argument and only uses the bound variable found within the slice, which is {\tt attributes}. Due to this reason, \oldtoolname{} can generate a correct update script for the second example (as shown in Figure~\ref{fig:old_coccievolve_result_correct}), but not for the first example.
% \jlx{I don't understand above, why \oldtoolname{} cannot generate a correct update script for the normalized first example? Why the incorrect script would miss line 27?}
% \sa{I think my example is not too clear and there were some mistakes in my description. \oldtoolname{} can only resolve to the first found value/variable that is located within the same method scope. Therefore, in the first example as above, {\tt parameterVariable0} can only be resolved into the bound variable {\tt attributes}. However, for a correct update, we will need to definition of the {\tt attributes} variable which is {\tt builder.build()} (line 21). Furthermore, we will also need the definition for the {\tt builder} object (line 20)}

\begin{figure}[t]
	\centering
	\scriptsize{
\begin{lstlisting}[language=diff,numbers=none]
@bottomupper_classname@
expression exp0, exp1;
identifier iden0, classIden;
@@
...
// Created Update Script
+ if (android.os.Build.VERSION.SDK_INT >=
+   android.os.Build.VERSION_CODES.LOLLIPOP) {
+   AudioAttributes newParameterVariable0 = attributes;
+   classIden.setAudioAttributes(newParameterVariable0);
+ } else {
...
+ }
\end{lstlisting}
		\caption{Incorrect update script for {\tt setAudioAttributes} API invocation generated by \toolname{} based on the first example in Figure~\ref{fig:arguments_form}}\label{fig:old_coccievolve_result_wrong}
	}
\end{figure}

\begin{figure}[t]
	\centering
	\scriptsize{
\begin{lstlisting}[language=diff,numbers=none]
@bottomupper_classname@
expression exp0, exp1;
identifier iden0, classIden;
@@
...
+ if (android.os.Build.VERSION.SDK_INT >=
+   android.os.Build.VERSION_CODES.LOLLIPOP) {
+   AudioAttributes newParameterVariable0 = new
+       AudioAttributes.Builder().build();
+   classIden.setAudioAttributes(newParameterVariable0);
+ } else {
...
+ }
\end{lstlisting}
		\caption{Correct update script for {\tt setAudioStreamType} method invocation generated by \toolname{} based on the second example in Figure~\ref{fig:arguments_form}}\label{fig:old_coccievolve_result_correct}
	}
\end{figure}

This problem severely limits the coding styles that are acceptable as examples for \oldtoolname{}, which subsequently limits its effectiveness. This problem also prevents \oldtoolname{} from being able to produce a working update script for examples which contain free variables.
% out-of-method variables.
% like the one shown in Figure~\ref{fig:out_of_method_boundaries} for {\tt request} variable.

To alleviate this problem, \toolname{} uses Data Flow Analysis (DFA) to resolve the values of expressions used as arguments in an API method invocation. This resolver should handle all possible forms of Java expressions. This data flow analysis is built to gather and predict a set of possible values
% Yes we also predict values in some way. For example, if we find an object, we might need to predict which constructor is used to make this object.
at any given point inside the code. Hence, this functionality is able to predict and resolve the correct replacement values for any expressions used in the API invocation arguments.

We made a custom lightweight DFA for this purpose by using the symbol resolver from Java Symbol Solver that is a part of Javaparser~\cite{javaparser}. This DFA conducts a bottom-up search from the bound variables or expressions used as the API method invocation arguments and expands the search scope until it finds the value or the method definition referred by the expressions, or until it explores the entire file. Using this approach, we can predict the value of free variables that are referred by the bound variables. Values and method invocations that are found by this analysis are used to replace the original expressions. These replacements are done to ensure that the expressions used as API invocation arguments are in the form of literal expressions, static class members, method invocations, or object creations.

The workflow for this DFA is shown in Figure~\ref{fig:data_flow_analysis}. The DFA receives as an input the expression to be resolved. This expression is used as API invocation argument and can be in the form of a method invocation expression, name expression, field access expression, and literal expression. Each form of expression will require a specific processing as given in the workflow diagram. To give a better understanding of this workflow, assume an example code provided in Figure~\ref{fig:update_code_example}.

% CHANGING THE FIGURE COUNTER HERE
\setcounter{figure}{8}

\begin{figure}[t]
	\centering
	\scriptsize{
\begin{lstlisting}[language=java,numbers=none,sensitive=true,columns=flexible,basicstyle=\ttfamily]
43 private int duration = 9;
44 private int frequency = 3;
45 public int amplitude = duration / frequency;
46 public VibrationEffect createVibration(int time, 
           int amplitude) {
47     return VibrationEffect.createOneShot(time, amplitude);
48 }
...
69 public void onCreate() {
70     if (android.os.Build.VERSION.SDK_INT >= 
                android.os.Build.VERSION_CODES.O) {
71         vibrator.vibrate(createVibration(3, amplitude));
72     } else {
73         vibrator.vibrate(50);
74     }
75 }

\end{lstlisting}
		\caption{Update example code for {\tt vibrate(long)} API}\label{fig:update_code_example}
	}
\end{figure}

In this update example, we can see that the updated {\tt vibrate} API used a method invocation expression as its argument (line 71). According to the workflow, we resolved the method definition, {\tt createVibration(...)} (line 46), and processed it using the copy method and class definition.
% defined in Section~\ref{copy_method_class}. 
Then, we resolve the scope of the method. 
%A method scope is an object or class used to invoke the method. 
However, since {\tt createVibration(...)} method is a public method that can be referenced
% In most cases (according to the APIs tested), they unfold to either a constants or method invocations. We can't really say on how often does it happen due to the limitation of the number of API and after-update example. If someone reassign the duration or frequency, then the after-update example is different. The data flow analysis works on a bottom-up approach, thus will resolve on the last assignment definition first before searching further up.
directly, no object or class is used in its invocation, thus resulting in no scope to resolve. Next, we resolve the arguments of the method invocation. The first argument, {\tt 3} is an integer literal expression, thus no replacement is needed. However, the second argument, {\tt amplitude} is a name expression, so we need to resolve its definition. Resolving this argument results in its definition which is {\tt duration / frequency} (line 45). Since this definition still contains expressions in the form of name expression, we further resolve their values recursively. From this process, we found the literal expressions of {\tt 9} (line 43) for duration and {\tt 3} (line 44) for frequency. These expressions are used to replace their values in the {\tt amplitude} variable definition in line 45 definition resulting in:

\begin{lstlisting}[language=java,numbers=none,sensitive=true,columns=flexible,basicstyle=\ttfamily]
45 public int amplitude = 9 / 3;
\end{lstlisting}

In the end, we replace this definition of {\tt amplitude} variable into the value used as the API invocation argument in line 71, resulting in this updated API:
\begin{lstlisting}[language=java,numbers=none,sensitive=true,columns=flexible,basicstyle=\ttfamily]
71 vibrator.vibrate(createVibration(3,
        9 / 3));
\end{lstlisting}

% CHANGING THE FIGURE COUNTER HERE
\setcounter{figure}{7}

\begin{figure}[t]
	\centering
	\captionsetup{belowskip=1.0pt,aboveskip=3.0pt}
	\includegraphics[width=1.0\linewidth]{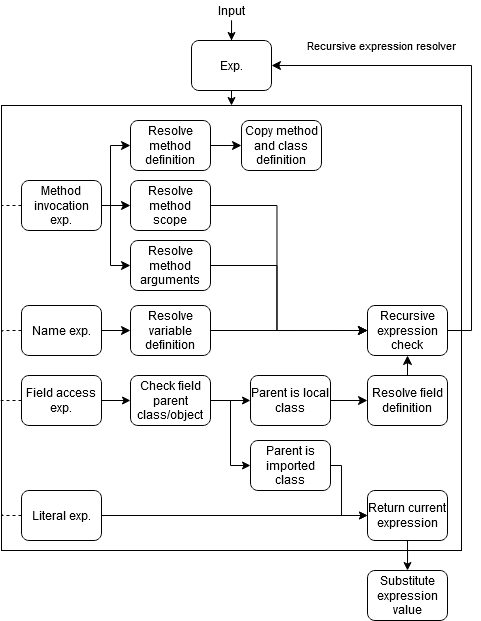}
	\caption{Overview of the data flow analysis workflow}
	\label{fig:data_flow_analysis}
\end{figure}

% CHANGING THE FIGURE COUNTER HERE
\setcounter{figure}{9}

This data flow analysis is built for \toolname{} as an upgrade from \oldtoolname{}. Therefore, the DFA is run as a preprocessing step before the update script creation. As a preprocessing step, this feature does not change the internal working behavior of the \oldtoolname{} but instead adds a layer of functionality that modifies the input code.

% \ft{What substitution actually happens here?}
% \sa{It substitute the original expressions with the expressions found through DFA. This is to ensure that the expressions used as API invocation arguments are the actual values and not just a variable that refers to other variables/class/method/etc. I think it need better wordings?}\ft{I don't really follow. Is it something like a=b; c=b; d.method(c); So c is original expressions and a is the actual value? If not, can you give a simple example?}\sa{yes, your illustration is correct. maybe to make it clearer:
% int a = 100; int b = a; someMethod(b); in this case, b is the original expression, and a is the actual value. It will be very helpful if you can help me explain this in clearer term as I don't really have the right definition and terms.}

\subsection{Source File Normalization}
Following the approach taken by \oldtoolname{}, Andro-Evolve also uses source file normalization in its workflow. Source file normalization is used to mitigate the problem of semantically-equivalent code being expressed in different forms, which can cause a failed API update. Variable normalization in \toolname{} is focused on the part of the file that contains the API invocations defined in the API update mapping, along with their arguments. Given an API invocation, source file normalization normalizes the code in three steps:

\begin{enumerate}[nosep,leftmargin=*]
    \item \toolname{} is the first Python API usage search tool that we know of;
    \item \toolname{} us
    \item Extract the returned value of the API invocation into variable assignment.
\end{enumerate}

Consider a code fragment containing a {\tt fromHtml(String)} API invocation as shown in the first example of Figure~\ref{fig:source_file_normalization}. Source file normalization will convert the code given in the first example into the normalized form given in the second example. First, all arguments, including the class or object used in the API invocation, are extracted. This extraction introduces the variables {\tt classNameVariable} (line 2), containing the class used in the API invocation, and {\tt parameterVariable0} (line 3), containing the argument of the API invocation. Next, the return value of the API invocation is also extracted, resulting in the {\tt tempFunctionReturnValue} variable (line 4--5).

\begin{figure}[t]
	\centering
	\scriptsize{
\begin{lstlisting}[language=java,numbers=none,sensitive=true,columns=flexible,basicstyle=\ttfamily]
// First Example
1 Spanned span = Html.fromHtml("<h2>Title</h2><br>");

// Second Example
2 Html classNameVariable = Html;
3 String parameterVariable0 = "<h2>Title</h2><br>";
4 Spanned tempFunctionReturnValue;
5 tempFunctionReturnValue = classNameVariable.
         fromHtml(parameterVariable0);
6 span = tempFunctionReturnValue;

\end{lstlisting}
		\caption{An illustration of the source file normalization result}\label{fig:source_file_normalization}
	}
\end{figure}

\subsection{Copying Method and Class Definition}
To provide a correct update, substituting the expressions into the resolved value is insufficient if the expressions are in the form of method invocation or object instantiation. This is because these expressions require their definitions to be used. Accordingly, we must copy method and class definitions from the resolved expression.

There are several important points to be considered for this feature. First, the copied class or method should be defined within the file containing the after-update example. This is due to \toolname{} limitation as a tool that works on a file scope. Therefore, if the class or method is defined outside of the after-update example file, \toolname{} will not be able to resolve them. 
Second, the copied class or method must be given an unique name, as required by Java.
% things to be considered is the uniqueness of the class or method name. Every class and every method within a class must have a unique name as Java does not allow duplicate name.
Lastly, the class or method that is copied must be in a scope that is accessible by the API invocation in the target file.

\begin{figure}[t]
	\centering
	\includegraphics[width=0.85\linewidth]{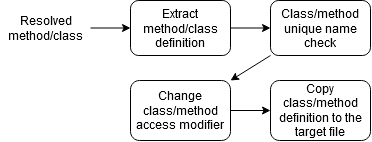}
	\caption{Overview of the copy method and class definition workflow}
	\label{fig:copy_method_class}
\end{figure}

The workflow of this feature can be seen in Figure~\ref{fig:copy_method_class}. First we extract the 
definitions of the methods and classes referenced by the expression from the code.
% expression method and class definition from the code.
Then, we make sure that in the target file, there are no class nor method with the same name as the extracted class and method. If a duplicate name is detected, the copied class or method name that is a duplicate is modified by adding number to the name. After validating the name, we then modify the access modifier of the class and method to {\tt public} to make sure that the API invocation in the target file can access them. After all these process, we then copy the class and method definition into the end of the target file.

Figure~\ref{fig:fig:method_copy_illus} illustrates this task.
Lines 1--7 show an example for {\tt requestAudioFocus(...)} API. The updated API uses an {\tt AudioFocusRequestOreo} object (line 3--4) as an argument but it is not resolved.
To correct the update, we must resolve and copy the relevant class definition, and instantiate the {\tt AudioFocusRequestOreo} object. Lines 10--51 show the results of this process. The {\tt AudioFocusRequestOreo} object is instantiated (line 11) and the relevant class is added into the updated code (line 30--51).

\begin{figure}[t]
	\centering
	\scriptsize{
\begin{lstlisting}[language=java,numbers=none,sensitive=true,columns=flexible,basicstyle=\ttfamily]

// Example of unresolved class audioFocusRequestOreo 
1  if (android.os.Build.VERSION.SDK_INT >= 
2          android.os.Build.VERSION_CODES.O) {
3      AudioFocusRequest request = audioFocusRequestOreo.
              getAudioFocusRequest();
4      result = audioManager.requestAudioFocus(request);
5  } else {
6      result = audioManager.requestAudioFocus(listener,
              type, duration);
7  }


// Example after resolving audioFocusRequestOreo class
10 if (android.os.Build.VERSION.SDK_INT >= 
         android.os.Build.VERSION_CODES.O) {
11     AudioFocusRequest request = new AudioFocusRequestOreo
               (this).getAudioFocusRequest();
12     result = audioManager.requestAudioFocus(request);
13 } else {
14     result = audioManager.requestAudioFocus(listener,
              type, duration);
15 }
    ...
30 class AudioFocusRequestOreo {
31     private AudioFocusRequest audioFocusRequest;
32     public AudioFocusRequestOreo(AudioManager.
33             OnAudioFocusChangeListener listener) {
        ...
40     }
41     public AudioFocusRequest getAudioFocusRequest() {
        ...
50     }
51 }


\end{lstlisting}
		\caption{Sample updates for  {\tt requestAudioFocus()} deprecated API}\label{fig:fig:method_copy_illus}
	}
\end{figure}

\subsection{Variable Name Denormalization}
Another problem of CocciEvolve is the temporary variables that are added during the update process. These temporary variables add multiple lines of code to the updated file which can be considered harmful since it makes the code less readable and understandable. Typically, most of the added lines just reference other variables in the code -- thus they are unnecessary. An example of the temporary variables created by CocciEvolve can be seen in Figure~\ref{fig:unnormalized_code}. In this figure, we can see that the temporary variables named {\tt parameterVariable} (line 11--16) and {\tt classNameVariable} (line 17) only refer to other variables that already exist in the file (method parameter in line 10).

\begin{figure}[t]
	\centering
	\scriptsize{
\begin{lstlisting}[language=java,numbers=none,sensitive=true,columns=flexible,basicstyle=\ttfamily]
    ...
10 public int saveLayer(float left, float top, float right, 
            float bottom, @Nullable Paint paint,
            int saveFlags) {
11 	float parameterVariable0 = left;
12 	float parameterVariable1 = top;
13	float parameterVariable2 = right;
14	float parameterVariable3 = bottom;
15	Paint parameterVariable4 = paint;
16	int parameterVariable5 = saveFlags;
17	Canvas classNameVariable = mCanvas;
18	if (VERSION.SDK_INT >= 21) {
19	    tempFunctionReturnValue = classNameVariable.
	        saveLayer(parameterVariable0, 
	        parameterVariable1, parameterVariable2, 
	        parameterVariable3, parameterVariable4);
20	} else {
21	    tempFunctionReturnValue = classNameVariable.
	        saveLayer(parameterVariable0,
	        parameterVariable1, parameterVariable2,
	        parameterVariable3, parameterVariable4,
	        parameterVariable5);
22	}
23 }
    ...
\end{lstlisting}
		\caption{Sample CocciEvolve update result for deprecated API {\tt saveLayer}}\label{fig:unnormalized_code}
	}
\end{figure}

Addressing this problem will be beneficial towards the readability and ease of understanding of the updated code. It will also make the code closer to developer-written code and less artificial. Variable name denormalization is our proposed approach for this purpose. Through the use of this denormalization, we aim to remove all unnecessary temporary variables and replace them with their values or referred variables. An example of the denormalized code based on the example from Figure~\ref{fig:unnormalized_code} can be seen in Figure~\ref{fig:denormalized_code}. In Figure~\ref{fig:denormalized_code}, temporary variables are removed and the relevant values are used directly in the method invocations. This results in a shorter, more concise, and more understandable code compared to the CocciEvolve update result.

\begin{figure}[t]
	\centering
	\scriptsize{
\begin{lstlisting}[language=java,numbers=none,sensitive=true,columns=flexible,basicstyle=\ttfamily]
    ...
10 public int saveLayer(float left, float top, float right,
        float bottom, @Nullable Paint paint, int saveFlags) {
11	if (VERSION.SDK_INT >= 21) {
12	    tempFunctionReturnValue = mCanvas.saveLayer(left, 
	        top, right, bottom, paint);
13	} else {
14	    tempFunctionReturnValue = mCanvas.saveLayer(left,
	        top, right,bottom, paint, saveFlags);
15	}
16 }
    ...
\end{lstlisting}
		\caption{Denormalized code of the one shown in Figure~\ref{fig:unnormalized_code}}\label{fig:denormalized_code}
	}
\end{figure}

\begin{figure}[t]
	\centering
	\captionsetup{belowskip=1.0pt,aboveskip=3.0pt}
	\includegraphics[width=0.95\linewidth]{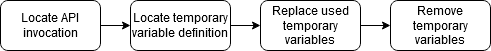}
	\caption{Overview of the variable denormalization workflow}
	\label{fig:variable_denorm}
\end{figure}

The steps in the denormalization process are shown 
% An illustration of how the denormalization work can be seen
in Figure~\ref{fig:variable_denorm}. First, we find the location of the API invocation that utilizes the temporary variables. For each of the temporary variable used in the API invocation, we resolve and locate its definition. These variable definitions contain their resolved values as the assigned expressions. These values are then used to replace the temporary variables used as the API's arguments. Finally, we delete the declaration and definition of the temporary variables as they are no longer needed.

\section{Experiment}\label{sec:exp}
\subsection{Dataset}\label{sec:dataset}
Our dataset contains 360 target files containing 20 Android deprecated-APIs, extended from the APIs that were used to evaluate \oldtoolname{}~\cite{coccievolve}. The dataset is obtained from a randomly selected GitHub projects obtained using AUSearch~\cite{asyrofi2020ausearch}, a tool to search Github repositories for API usages. Using this tool, we collected public GitHub repositories that contain invocations of the deprecated APIs and their replacement APIs in our dataset. Our dataset comprises after-update examples, target files to update that contains usages of the APIs, and one-to-one mappings from the deprecated APIs to the replacement APIs. %The mapping describes the update to be done based on the after-update example code and the target code. 
% \ft{I think after-update examples and target code are explained in the earlier part of the paper so we don't need to explain them again here?}\ft{Don't we also need the mapping from deprecated API to replacement API? We also only handle one-to-one mapping right.}\sa{maybe just delete the explanation of target code and after-update example from this paragraph? I have added the statement about the mapping}
% After-update examples are used to learn update scripts. This example contains both the deprecated API method and the updated API method in the form of an if code block. An example of this update example can be seen in Figure~\ref{fig:after_update_example}. Target apps are Android applications that contains the deprecated API which need to be updated.
Detailed statistics of the target files are shown in Table~\ref{table:data_statistic}.
% \ft{We need to make tables and figures to be in the same page they are referred to. We can fix this last.}\sa{I think we need to fix this later. However I don't really know how to.}\ft{I usually just move the tables/figures to another lines and see whether they appear in the same page.}

\begin{table}[t]
\caption{Number of targets in our evaluation dataset}
\begin{center}
\begin{tabular}{ |p{20em}|p{4.5em}| }
\hline
 \textbf{API Description} & \textbf{\# Targets}
\\ \hline
addAction(...) & 2
\\ \hline
getAllNetworkInfo() & 11
\\ \hline
getCurrentHour() & 60
\\ \hline
getCurrentMinute() & 60
\\ \hline
setCurrentHour(Integer) & 32
\\ \hline
setCurrentMinute(Integer) & 15
\\ \hline
setTextAppearance(...) & 15
\\ \hline
addGpsStatusListener(...) & 10
\\ \hline
fromHtml(...) & 15
\\ \hline
release() & 11
\\ \hline
removeGpsStatusListener(...) & 5
\\ \hline
shouldOverrideUrlLoading(...) & 0
\\ \hline
startDrag(...) & 4
\\ \hline
abandonAudioFocus(...) & 1
\\ \hline
getDeviceId() & 29
\\ \hline
requestAudioFocus(...) & 53
\\ \hline
saveLayer(...) & 21
\\ \hline
setAudioStreamType(...) & 2
\\ \hline
vibrate(long) & 8
\\ \hline
vibrate(long[], int) & 6
\\ \hline

\end{tabular}
\end{center}
\label{table:data_statistic}
%\vspace{-0.2cm}
\end{table}

As shown in Table~\ref{table:data_statistic}, there are no target files for the API {\tt shouldOverrideUrlLoading(...)}. During the search, we did not find any after-update example nor target file for it.

\subsection{Research Questions}\label{sec:rq}
%We conduct two different experiments on the available dataset, performance measure and readability measure.

\subsubsection{RQ1: How many updates can \toolname{} apply correctly?}
\hfill \break
%The first experiment is performance measure, in which we measure the performance of \toolname. 
We assess update performance by counting the number of correct updates produced. A correct update is an update that contains the deprecated and replacement API method in the form of an {\tt if} code block, alongside with all the methods and classes needed by the replacement API method.
% \ft{Is this a correct definition? Applicable should mean that we can apply the example to the target code while correct means that the applied update is correct (the update can possibly be applied yet incorrect.)} \sa{I think we remove the applicable term? Just leaving the correct update}\ft{Yes, I think we can do that for this paper.}
We compare the update performance of \toolname{} and CocciEvolve. %For both \toolname{} and CocciEvolve, we use a single update example for each API. This single example is taken from the updated API example provided in the AppEvolve replication package. 
%We then count the amount of correct and applicable updates for each tools and compare their performance. 
We also ask an experienced Java and Android engineer to check the measured update performance and verify their correctness. %The updated codes obtained through this experiment are then used for our second experiment, readability measure.

\subsubsection{RQ2: How readable is the updated code produced by \oldtoolname{} and \toolname{}?}
\hfill \break
We measured the readability of the updated code produced by \toolname{} and CocciEvolve. %These updated codes are the result of our first experiment. 
In order to get a better insight on the readability aspect of the update, we conduct an automatic and a manual measurement. 

In the automatic measurement, we utilized a state-of-the-art code readability scoring tool proposed by Scalabrino et al.~\cite{readabilitymodel}. %This research is one of the most recent state of the art study on the estimation of code readability. The model used in this readability estimation is comprised of both structural and textual features of the code. Structural features include the length of each line, indentation, usage of spaces, number of keywords, and other structural features. Textual features are based on the words used in the code, including their consistency, the usage of terms from dictionary, the meaning of the words used, and textual coherence. 
The tool outputs a code readability score in a scale of 0.0 to 1.0 with the higher score being a measure of better readability. As the readability score is affected by the length of the source code file, we performed a static slicing to obtain parts of the code that are affected by the update. 
% An illustration of the this static slicing is shown in Figures~\ref{fig:readability_slicing}.\ft{The figure may not be necessary.}\sa{can be deleted i think. It isn't an important part of \toolname{}}
The static code slicing is done using JavaParser\cite{javaparser} by first locating the deprecated and updated Android APIs based on their description. After these APIs are found, we slice the API method invocations and all the variables that are used in the invocations. The sliced code is then put into a template class and method to allow readability measurement using the tool. An example of the sliced code file is shown in Figure~\ref{fig:code_slicing_example}.

%However, we noticed that the codes varies greatly, thus resulting in wildly different readability measure depending on the file. To mitigate this problem, we decided to do a simple static slicing to obtain the parts of the code that are affected by the update. An illustration of this static slicing can be seen in Figure~\ref{fig:readability_slicing}. 

% \begin{figure}[t]
% 	\centering
% 	\includegraphics[width=0.95\linewidth]{Diagrams/Readability Slicing.png}
% 	\caption{Overview of the readability measure static slicing process}
% 	\label{fig:readability_slicing}
% \end{figure}

\begin{figure}[t]
	\centering
	\scriptsize{
\begin{lstlisting}[language=java,numbers=none,sensitive=true,columns=flexible,basicstyle=\ttfamily]
class MainActivity {
	public static void main() {
		int parameterVariable0 = lastHour + 1;
		TimePicker classNameVariable = timePicker;
		if (Build.VERSION.SDK_INT >= 
		        Build.VERSION_CODES.M) {
		    classNameVariable.setHour(parameterVariable0);
		} else {
		    classNameVariable.setCurrentHour(
		            parameterVariable0);
		}
	}
}
\end{lstlisting}
		\caption{Sliced code example for deprecated {\tt setCurrentHour(...)} API}\label{fig:code_slicing_example}
	}
\end{figure}

In the manual measurement, we asked two experienced Android developers to score 60 updated code, with 30 updated code each from CocciEvolve and \toolname. We choose 30 as a sample size to represent the variation in the syntax of the updated code. The developers were not told about which tool is used to produce the updated code. The first developer has 5 years experience in Android, while the second developer has 3 years of experience. For each updated code, the developers were asked to give score in the Likert scale of 1-5 for the readability of the code, and the naturalness of the code. A higher score indicates higher readability and higher confidence that the code resembles the one produced by human. For each pair of updated code produced by CocciEvolve and \toolname, the developers are also asked to determine which code that they prefer.

\subsubsection{RQ3: How efficient is \toolname{} in producing updates?}
\hfill \break
We measured the time needed for \toolname{} to update the target file. Specifically, we measure the time to perform update script creation and update application. The measurement is conducted in a Macbook Pro with a 2.3 GHz Intel Core i5 processor, and 8 GB 2133 MHz random access memory. The system ran Java SE 11, with OpenJDK version 11.0.6.

%Given an after-update example and its corresponding API update mapping, update script creation process creates the update script for the API described in the mapping. In the update application process, we apply the update script into the specified target code based on the API update mapping. \ft{already explained in approach}For each of process, we calculate the time needed by \toolname{} before producing any result.
%We measured the time needed for \toolname{} to update the target code. For this measurement, we separated the process within \toolname{} into two: update script creation process, and update application process. Given an after-update example and its corresponding API update mapping, update script creation process creates the update script for the API described in the mapping. In the update application process, we apply the update script into the specified target code based on the API update mapping. For each of process, we calculate the time needed by \toolname{} before producing any result.

\subsection{Results}\label{result_subsection}
%We conduct the experiments based on the experiment settings detailed in Section \ref{experiment_settings}. The experiments consist of performance measure that measure the amount of correct update produced, and readability measure that estimate the code readability improvement of \toolname.

\subsubsection{RQ1: Code Update Accuracy}\label{sec:RQ1}
%\hfill \break
%In this experiment, we added improvement in the form of data flow analysis into \toolname{}. Data flow analysis allowed \toolname to resolve the correct values of the new API invocation arguments that are introduced in the updated API. 
In total, \toolname{} and \oldtoolname{} managed to correctly update 316 and 249 out of the 360 target files, respectively. AndroEvolve outperforms CocciEvolve by 26.90\%. %from the 16 tested APIs.
Analysis on the results shows that the inclusion of data flow analysis improves the update result of \toolname{} significantly, specifically in the {\tt vibrate(long)}, {\tt vibrate(long[],int)}, and {\tt requestAudioFocus(...)} APIs. After-update examples for these APIs include usages of out-of-method variables that are not handled by \oldtoolname{}. \toolname{} has similar performance as \oldtoolname{} for APIs that do not use out-of-method variables in the after-update example.

% NOTES: maybe add example of the examples that contain non-closed/non-resolved values. Or maybe can add this in introduction or approach (DFA part)

Despite the big improvement when compared to CocciEvolve, \toolname{} still has several problems. One of these problems affects {\tt addGpsStatusListener(...)} and {\tt removeGpsStatusListener(...)} APIs. These APIs utilize out-of-file variables for their method invocation. Out-of-file variables are variables that are defined outside of a file scope. These variables can be a class member or a method invocation argument that are defined in other files. Since \toolname{} works only in a file scope, these out-of-method variables are unresolved, causing an incomplete update. An example of the updated code that carries this problem is shown in Figure~\ref{fig:input_parameter}. In this example, the {\tt callback} object (line 2, 6) is a class member whose value is defined in another file.

\begin{figure}[t]
	\centering
	\scriptsize{
\begin{lstlisting}[language=java,numbers=none,sensitive=true,columns=flexible,basicstyle=\ttfamily]
1  public class MixedPositionProvider extends PositionProvider
        implements LocationListener, GpsStatus.Listener {
2      private GnssStatus.Callback callback;
3      public void startUpdates() {
4      GpsStatus.Listener listener = this;
5          if (android.os.Build.VERSION.SDK_INT >=
                android.os.Build.VERSION_CODES.N) {
6               locationManager.registerGnssStatusCallback(
                    callback);
7          }
8          else{
9             locationManager.addGpsStatusListener(listener);
10         }
11     }
12 }
\end{lstlisting}
		\caption{An example usage of out-of-file variable in {\tt addGpsStatusListener} method invocation}\label{fig:input_parameter}
	}
\end{figure}

\toolname{} cannot handle complex updates that involve an update of a single API into multiple APIs, such as updating {\tt getAllNetworkInfo()} API to {\tt getAllNetworks()} and {\tt getNetworkInfo(...)} APIs. Unlike the usual Android API update that replaces an API with a newer API, the case of {\tt getAllNetworkInfo()} is different. Updating this API involves the addition of a new control flow in the form of a loop that iterates the {\tt Network} object returned by {\tt getAllNetworks()} to receive their {\tt NetworkInfo} object by using {\tt getNetworkInfo(...)} method.
%\ft{This is vague.} \sa{added more explanations}
%This feature is not yet supported in the current version of \toolname{}. 

%Apart from these problematic APIs, \toolname{} only sees minor drawback. The only problem that is found outside of the problematic APIs is on cases in which the same API method is invoked multiple time in a single line of code.
\toolname{} also cannot update multiple invocations of an API method written in a single line of code.
While uncommon, this problem has been found in some target files, resulting in an incomplete update. An example of this problem is shown in Figure~\ref{fig:multiple_invocation}. We can see that {\tt getCurrentHour()} (line 7) is invoked multiple times in the last line, causing only the first invocation to be updated.

\begin{figure}[t]
	\centering
	\scriptsize{
\begin{lstlisting}[language=java,numbers=none,sensitive=true,columns=flexible,basicstyle=\ttfamily]
1 int tempFunctionReturnValue;
2 if (android.os.Build.VERSION.SDK_INT >= 23) {
3     tempFunctionReturnValue = timePickerBegin.getHour();
4 } else {
5     tempFunctionReturnValue = timePickerBegin.
            getCurrentHour();
6 }
7 dateTime = tempFunctionReturnValue + ":" + 
        timePickerBegin.getCurrentMinute() + "-" + 
        timePickerEnd.getCurrentHour() + ":";
\end{lstlisting}
		\caption{An example of multiple invocations of {\tt getCurrentHour()} method in a single line}\label{fig:multiple_invocation}
	}
\end{figure}

\subsubsection{RQ2: Code Readability}
%\hfill \break
%For the readability measure, we add the variable name denormalization into \toolname{}. This additional features allow \toolname{} to remove the temporary variables that were present in the \oldtoolname{} update and substitute them with their original value from the target code. Measurement of this readability is done through automatic scoring using tools provided from the "A Comprehensive Model for Code Readability" paper~\cite{readabilitymodel}, and manual scoring on the codes readability and human-likeness from two experienced Android engineers that are not part of this project.

%In the automatic scoring, we used all the results from the performance measure and utilized code slicing to obtain the all the codes that are related to the API update. 

In automatic readability measurement, we compute the scores of all 355 updated code and average the scores for the APIs. %Scores measured by tools are in the range of 0.0 to 1.0, with higher score being a better readability. 
The detailed scores are shown in Table~\ref{table:readability_measure}. Based on this result, code updated by \toolname{} has higher scores for all APIs. Further analysis of the updated code shows that a bigger improvement is observed for APIs with multiple arguments (e.g. {\tt saveLayer(...), startDrag(...)}, etc.). 
%On the contrary, for APIs that do not have any argument, we observe less improvement in readability scores.

\begin{table}[t]
  \centering
  \caption{Updated code automated readability scores}
  \label{table:readability_measure}
  \begin{tabular}{|l|c|c|c|}
 
 \hline
\multicolumn{1}{|c|}{\multirow{2}{*}{\textbf{API}}} & \multicolumn{1}{|c|}{\multirow{2}{*}{\textbf{\# Code}}}  & \multicolumn{2}{|c|}{\textbf{Average Score}} \\ 
\cline{3-4} 
\multicolumn{1}{|c|}{} & \multicolumn{1}{|c|}{} &
\textbf{AndroEv}           & \textbf{CocciEv}          \\ \hline
 
addAction(...) & 0 & 0 & 0
\\ \hline 
getAllNetworkInfo()  & 11 & 0 & 0
\\ \hline
getCurrentHour()  & 60 & 0.6009 & 0.5071 
\\ \hline
getCurrentMinute() & 60 & 0.5996 & 0.5172
\\ \hline
setCurrentHour(Integer)  & 32 & 0.6006 & 0.3904 
\\ \hline
setCurrentMinute(Integer)  & 15 & 0.5766 & 0.3962
\\ \hline
setTextAppearance(...) & 15 & 0.5615 & 0.2947
\\ \hline
addGpsStatusListener(...) & 10 & 0.3945 & 0.2307 
\\ \hline 
fromHtml(...) & 15 & 0.4143 & 0.2593
\\ \hline 
release() & 11 & 0.8311 & 0.6890
\\ \hline
removeGpsStatusListener(...) & 5 & 0.4006 & 0.2287
\\ \hline
shouldOverrideUrlLoading(...) & 0 & 0 & 0 
\\ \hline
startDrag(...) & 4 & 0.4516 & 0.1440
\\ \hline
abandonAudioFocus(...) & 0 & 0 & 0
\\ \hline 
getDeviceId() & 29 & 0.4545 & 0.3974
\\ \hline
requestAudioFocus(...) & 53 & 0.2413 & 0.2290
\\ \hline 
saveLayer(...)  & 21 & 0.4115 & 0.1011
\\ \hline
setAudioStreamType(...) & 0 & 0 & 0 
\\ \hline
vibrate(long) & 8 & 0.5284 & 0.3629
\\ \hline 
vibrate(long[], int) & 6 & 0.4437 & 0.2631 
\\ \hline 
  \end{tabular}
  
\end{table}

%To further strengthen our finding, we conducted manual readability measure by taking random samples of the updated code from both \toolname{} and CocciEvolve and asked two experienced Android developers to score them. The developers were asked to give a score of 1-5 on the code readability and the human-likeness of the code. In total, each developers give scores to 60 updated codes, which are a pair of 30 CocciEvolve updated codes, and 30 \toolname{} denormalized updated codes. Matching the result from the automated scoring, manual scoring also results in the denormalized code scored better than the CocciEvolve updated code.

The manual readability measurement strengthens the above findings. The average readability score given by the developers for code updated by \toolname{} is 4.817. Meanwhile, the average readability score for code updated by \oldtoolname{} is 2.633. Improvement can also be seen in the scores for code naturalness:
\toolname{}'s code was given an average score of 4.917, while \oldtoolname{}'s only received an average score of 2.433 for the naturalness aspect of their code.

\subsubsection{RQ3: Update time}
%We measure the time needed by \toolname{} to perform each of its process, namely update script creation process and update application process. For update script creation process, we measure the time taken to produce the update script using \toolname{}. For update application process, we measure the time taken for the application of update script into the target code.\ft{already explained in RQ}
The update time of \toolname{} is shown in Table~\ref{table:time_measurement}.
%The result of this measurement can be seen in Figure~\ref{table:time_measurement}.

\begin{table}[t]
\caption{Time measurement results of \toolname{} (in seconds)}
\begin{center}
\begin{tabular}{ |p{12em}|p{5em}|p{5em}| }
\hline
 \textbf{API} & \textbf{Update Creation} & \textbf{Update Application}
\\ \hline
addAction(...) & 9.601 & -
\\ \hline
getAllNetworkInfo() & 9.083 & 9.417
\\ \hline
getCurrentHour() & 13.840 & 11.140
\\ \hline
getCurrentMinute() & 8.660 & 10.172
\\ \hline
setCurrentHour(Integer) & 6.365 & 7.133
\\ \hline
setCurrentMinute(Integer) & 9.345 & 6.173
\\ \hline
setTextAppearance(...) & 7.935 & 13.144
\\ \hline
addGpsStatusListener(...) & 9.434 & 6.206
\\ \hline
fromHtml(...) & 9.278 & 13.448
\\ \hline
release() & 10.792 & 6.622
\\ \hline
removeGpsStatusListener(...) & 9.316 & 6.500
\\ \hline
shouldOverrideUrlLoading(...) & - & -
\\ \hline
startDrag(...) & 10.030 & 6.483
\\ \hline
abandonAudioFocus(...) & 12.230 & -
\\ \hline
getDeviceId() & 9.231 & 9.001
\\ \hline
requestAudioFocus(...) & 9.340 & 11.112
\\ \hline
saveLayer(...) & 8.767 & 6.589
\\ \hline
setAudioStreamType(...) & 9.197 & -
\\ \hline
vibrate(long) & 8.700 & 5.876
\\ \hline
vibrate(long[], int) & 8.292 & 5.747
\\ \hline

\end{tabular}
\end{center}
\label{table:time_measurement}
%\vspace{-0.2cm}
\end{table}

Both update-script creation and update-script application steps in \toolname{} took an average of less than 15 seconds to execute. Given an after-update code example, a target file, and an API update mapping, \toolname{} can update the API usages in the target file in less than a minute.
%This shows the efficiency of \toolname{} in updating deprecated Android API usages.

%Measurement of time taken for update creation and update application processes shows that on average, \toolname{} took less than 15 seconds for each process. Given an after-update code example, a target code, and API update mappings, \toolname{} will be able to update the API usage of the target code in less than a minute. This shows the efficiency of the usages of \toolname{} for updating API usages in Android deprecated APIs.

Further analysis yields several observations. First, aside from API complexities, the number of API invocations in the target file also affects the update time. Second, when processing multiple files containing the same API usage, the time can be shortened by using the same update script.
%This finding suggests that \toolname{} is suitable to process multiple codes in a project environment.

%Analysis during the time measurement also yields several findings. First, aside from the different complexities between APIs, the number of API invocations present in the code also affect the time needed by \toolname{}. Some target code may contain several of the same API invocations, affecting the time execution of the update application. Second finding is found when processing multiple files using \toolname{}. By updating multiple files containing the same API usage, we can minimize the time execution of \toolname{} through the use of the same update script for the update. This finding suggests that \toolname{} is suitable to process multiple codes in a project environment.

\section{Discussion}\label{sec:discuss}
Based on the results provided in Section~\ref{result_subsection}, it is evident that \toolname{} achieves a better performance than \oldtoolname{}. Our approach solves the problem of out-of-method variables by using data flow analysis to resolve their values. Moreover, variable name denormalization also improves the readability of the updated code, as demonstrated by our automatic and manual readability measurements. %We evaluated \toolname{} in a dataset of 355 target files from which 316 successful update is produced, compared to \oldtoolname{} that only managed to produce 249 successful update with worse readability aspect. This dataset contains code with 16 different Android APIs usage.

%However, this achievement leave the question regarding the update failures that happen within \toolname{}.
% \ft{The problems are redundant with Section~\ref{result_subsection}. Can we discuss some possibilities to fix the problems?} \sa{I will try to add the possibilities to fix in some sentences} \sa{added some sentences}
Despite this achievement, \toolname{} still fails to update some target files, as described in Section~\ref{sec:RQ1}. Some failures are caused by multiple invocations of the same API within a single line of code. A possible solution for this problem is to temporarily separate the same API invocations that exist in the same line of code into multiple statements to be updated independently. Other failures occur due to usages of out-of-file-boundary variables in the after-update example. To mitigate this problem, we need to improve the data flow analysis to make it work in a project scope, and change the input of \toolname{} from a file into a project. Problems also occur in the case of APIs with complex updates that convert a single API invocation into multiple API invocations. Solving this problem will require an overall change in \toolname{}, by allowing it to accept a non-one-to-one API update mapping.

\begin{figure}[t]
	\centering
	\scriptsize{
\begin{lstlisting}[language=java,numbers=none,sensitive=true,columns=flexible,basicstyle=\ttfamily]
1 public ChromeNotificationBuilder addAction(int icon, 
        CharSequence title, PendingIntent intent) {
2         if (Build.VERSION.SDK_INT >= Build.VERSION_CODES.M) {
3             Notification.Action action = new Notification.
                Action.Builder(Icon.createWithResource
                (mContext, icon), title, intent).build();
4             mBuilder.addAction(action);
5         } else {
6             mBuilder.addAction(icon, title, intent);
7         }
8         return this;
9     }
\end{lstlisting}
		\caption{Code example for {\tt } API that use inner class constructor}\label{fig:inner_class_constructor}
	}
\end{figure}

Another problem occurs when the after-update example makes use of an inner class constructor. This problem occurs due to a limitation of Coccinelle4J, which \toolname{} inherits. Coccinelle4J only supports middleweight Java, which only includes a subset of the Java grammar. This subset does not include the inner class constructor. As a consequence, \toolname{} cannot handle an after-update example that contains inner class constructors. Such an example is shown in Figure~\ref{fig:inner_class_constructor}. In this example, the creation of the {\tt action} object (line 3) uses an inner class constructor {\tt Notification.Action.Builder}. This problem affects the update of several APIs, including {\tt addAction(...)}, {\tt abandonAudioFocus(...)}, and {\tt setAudioStreamType(...)}. %After-update code example for these APIs contains inner class constructors.

\section{Related Work}\label{sec:related}
% \ft{We also need to talk about AppEvolve and CocciEvolve.} \sa{will add discussion and citation to them} \sa{I add discussion about them in the end of the third paragraph}
\textbf{API deprecation}.
Studies about API deprecation have been done frequently \cite{li2018characterising, zhou2016api, brito2016developers, yang2018android, robbes2012developers, horadevelopersapievolution, sawant2018reaction, fazzini2019automated, coccievolve}.
Li et al.\cite{li2018characterising} proposed a tool called CDA to characterize deprecated Android APIs. They found inconsistent annotation and documentation on deprecated APIs, and that most deprecated APIs are used in popular libraries.
Zhou et al.\cite{zhou2016api} examined the usages of deprecated APIs in 26 open-source Java frameworks and libraries and found that many of these APIs were never updated. They proposed Deprecation Watcher, a tool to detect deprecated Android API usages from code examples on the web.
Brito et al.\cite{brito2016developers} conducted a large scale analysis on Java systems to measure the usages of deprecation messages. Their analysis showed that a number of deprecated APIs did not use these replacement messages.
Yang et al.\cite{yang2018android} investigated the impact of Android OS updates on Android apps. They presented an automatic approach to detect parts of Android apps affected by an OS update.
% \ft{Is this work about deprecation or any type of update?} \sa{This work is about the deprecation that is caused by the operating system update, referring to the Android APIs that become obsolete and affecting some applications}

%Studies about API deprecation have been done frequently \cite{li2018characterising, zhou2016api, brito2016developers, yang2018android, robbes2012developers, horadevelopersapievolution, sawant2018reaction}. Li et al.\cite{li2018characterising} proposed a tool called CDA to characterize deprecated Android APIs. Their investigation result in several findings, namely inconsistent annotation and documentation of deprecated API, and that most usage of deprecated APIs are located in popular libraries. Zhou et al.\cite{zhou2016api} examined the usages of API deprecation in 26 open source Java frameworks and libraries, finding that a lot of deprecated APIs are not updated. They then proposed Deprecation Watcher, a lightweight tool to detect deprecated Android API usages from code examples on the web. Brito et al.\cite{brito2016developers} conducted a large scale analysis on Java systems to measure the usages of deprecation messages. Their analysis show that there are deprecated APIs that did not use these replacement messages. Study by Yang et al.\cite{yang2018android} investigates the impact of Android operating system update towards applications. They present an automatic approach to detect parts of Android apps that are affected by the update.

Some studies focus on the effect of API deprecation.\cite{robbes2012developers, horadevelopersapievolution, sawant2018reaction}. Robbes et al.\cite{robbes2012developers} conducted a case study on the Smalltalk ecosystem and found that API deprecation messages are not always helpful. Hora et al.\cite{horadevelopersapievolution} conducted a case study on the Pharo ecosystem on the impact of API evolution, resulting in similar findings that API changes can have a large impact in the client systems, methods, and developers. They also found that API replacements can not be resolved uniformly.
% \ft{Does this work talk about API deprecation or API evolution in general?} \sa{In general, this work is more focused in the effect of API deprecation that is caused by the evolution. I added this reference since this is also a kind of replication study with the previous study by Robbes et al. on Smalltalk ecosystem (first reference in this paragraph)}
Sawant et al.\cite{sawant2018reaction} replicated the study on Java. They found that only a small fraction of developers react to API deprecation and most of these developers prefer to remove usages of deprecated APIs. 

Our study also deals with API deprecation. It focuses on automatically updating the usages of deprecated Android APIs. Similar studies on this topic have been done recently. AppEvolve\cite{fazzini2019automated} is one of the first tools proposed for this purpose. It performs API updates by learning from both before- and after- update example. 
% AppEvolve has four main steps: analyze the target file to identify code affected by the API changes; search existing code bases for updates examples; analyze, rank, and transform update examples into generic patches; apply the generated patches to the target file. 
\oldtoolname{}\cite{coccievolve} is the current state-of-the-art tool for automated update of deprecated Android API usage. \oldtoolname{} improves on AppEvolve by only using a single after-update example to perform API update and providing a highly readable and configurable update API update script in the form of semantic patches. \oldtoolname{} also solves the problem of failure to update code with different form or syntax that occurs in AppEvolve, as highlighted by the replication study by Thung et al.\cite{thung2020automated}. \oldtoolname{} utilizes parameter variable normalization to mitigate this problem.

%Some studies are focused on the effect of API deprecation.\cite{robbes2012developers, horadevelopersapievolution, sawant2018reaction}. Robbes et al.\cite{robbes2012developers} used a case study of Smalltalk ecosystem, finding that API deprecations can have a very large impact on the ecosystem and that deprecation messages are not always helpful. Hora et al.\cite{horadevelopersapievolution} conducted a case study in Pharo ecosystem on the impact of API evolution, resulting in similar findings that API changes can have a large impact in the client systems, methods, and developers. They also found that replacement of APIs can not be resolved in a uniform manner. Sawant et al.\cite{sawant2018reaction} proposed a partial replication study of the previous research through a study on Java language. Their study found that only a small fractions of developers react to API deprecation and most developers prefer to delete the call to the deprecated APIs. While our study is in the aspect of API deprecation, it is focused on automatically updating the usages of deprecated API in Android operating system.

\textbf{Program transformation}. Program transformations have been studied extensively \cite{Visser:2001:SLP:647200.718711, LASE, Meng:2013:LLA:2486788.2486855, Rolim:2017:LSP:3097368.3097417, lawall2018coccinelle, brunel2009foundation, kang2019semantic}.
Stratego\cite{Visser:2001:SLP:647200.718711} is a language for program transformation based on the paradigm of rewriting strategies. Stratego performed transformation following the written transformation rules. LASE\cite{LASE, Meng:2013:LLA:2486788.2486855} is an example based program transformation tool that is capable of locating and applying systematic edits. LASE provides users with a view of the syntactic edit and its corresponding contexts, allowing users to review and correct the edit suggestions. Rolim et al.\cite{Rolim:2017:LSP:3097368.3097417} proposed REFAZER, a technique for automatically learn program transformations by observing code edits performed by developers.

%Program transformation is a domain that have been studied extensively\cite{Visser:2001:SLP:647200.718711, LASE, Meng:2013:LLA:2486788.2486855, Rolim:2017:LSP:3097368.3097417, lawall2018coccinelle, brunel2009foundation, kang2019automating}. Multiple tools and frameworks have been proposed for program transformation. Stratego\cite{Visser:2001:SLP:647200.718711} is a language for program transformation based on the paradigm of rewriting strategies. Stratego performed transformation based on the transformation rules given as input. LASE\cite{LASE, Meng:2013:LLA:2486788.2486855} is an example based program transformation tool that is capable of locating and applying systematic edits. LASE provides users with a view of the syntactic edit and its corresponding contexts, allowing users to review and correct the edit suggestions. Rolim et al.\cite{Rolim:2017:LSP:3097368.3097417} proposed REFAZER, a technique for automatically learning program transformations. REFAZER observes code edits performed by developers and used such observations as input-output examples to learn program transformations.

Coccinelle\cite{lawall2018coccinelle, brunel2009foundation} is a C-based program matching and transformation tool that has been utilized for the automated evolution of Linux kernel. Coccinelle allows developers to write their transformation rules using Semantic Patch Language (SmPL). Recently, Kang et al.~\cite{kang2019semantic} proposed Coccinelle4J, a port of Coccinelle for Java language. Coccinelle4J allows the transformation of Java program using the same method as Coccinelle, through the use of semantic patch written in Semantic Patch Language. 

In our work, \toolname{} applies program transformation to update deprecated Android API usages. It uses SmPL to write the transformation and Coccinelle4J to apply it.

\section{Conclusion and Future Work}\label{sec:conclusion}
Updating the usages of deprecated Android APIs is a priority to ensure the functionality of Android apps in the current and previous versions of Android OS. However, performing such updates is time-consuming and labor-intensive. In this work, we proposed \toolname{}, an automated Android API usage update tool. \toolname{} uses data flow analysis to resolve the values of out-of-method variables, allowing \toolname{} to work on the file scope. \toolname{} also performs variable denormalization to produce updated code with good readability. We evaluated \toolname{} using a dataset of 360 target files from which it managed to produce 316 successful updates. On the same dataset, \oldtoolname{}, the previous state-of-the-art tool, only managed to produce 249 successful updates. We also evaluate the updated code readability using both manual and automatic measurements. In the manual measurement, we asked the opinions of two developers on the readability of the updated code, while in the automatic measurement, we used a code readability scoring tool. In both measurements, \toolname{} outperforms CocciEvolve by 49.89\% and 82.94\% respectively.

%Updating the usages of deprecated Android APIs is a priority to ensures the workability of the application in the current and previous versions of Android operating system. However, this work is time-consuming and labor-intensive for the developers due to the possible widespread of API usages. In this work, we proposed \toolname{}, an automated Android API usages update tool. \toolname{} features data flow analysis capability to resolve the values of out-of-method boundaries variables, allowing \toolname{} to work on a file scope. \toolname{} also produced human-like updated code with good readability by removing and substituting the usage of temporary variables introduced by the update. We evaluated \toolname{} in a dataset of 355 target codes from which it managed to produce 316 successful update. Meanwhile, previous state-of-the-art tool, \oldtoolname{} only managed to produce 249 successful update. Evaluation for updated code readability are also done using both manual and automatic approach. In manual approach, we asked the opinions of two engineers on the readability of the updated code, while in automatic approach, we used automated code readability scoring tools on the updated code. In both evaluation, \toolname{} achieves better overall code readability compared to \oldtoolname{}.

For future work, we plan to increase the capability of \toolname{}. First, we plan to improve the data flow analysis to allow resolving values that are located in other files within the same project. This addition will make \toolname{} capable of handling out-of-file variables. Second, we also plan to handle more complex Android API updates, especially for cases where a single API is updated into several different APIs. While such a case is uncommon, this improvement would increase the overall effectiveness of \toolname{}.

\balance
\bibliography{references}
\bibliographystyle{plain}

\end{document}